\documentstyle[12pt]{article}

\begin{document}
{\sf \begin{center} \noindent {\Large\bf Riemann curvature-stretching coupling in dynamo torus laboratory and in UHF twisted plasma loops}\\[1mm]
{\sl  L.C. Garcia de Andrade}\\Departamento de F\'{\i}sica
Te\'orica -- IF -- Universidade do Estado do Rio de Janeiro-UERJ\\[-2mm]
Rua S\~ao Francisco Xavier, 524\\[-2mm]
Cep 20550-003, Maracan\~a, Rio de Janeiro, RJ, Brasil\\[-2mm]
\end{center}
\paragraph*{}
 Kleeorin, Rogachevskii, Tomin and Sokoloff [Phys Rev \textbf{E79},046302,(2009)] have shown that Roberts slow dynamo can be transformed into a fast dynamo by allowing its coefficients randomly fluctuate. Stretching, fundamental for fast dynamo action, is increased by mean helicity flow. Previous investigation on slow dynamo plasma and anti-fast dynamo theorem in Riemannian geometry [Garcia de Andrade, Phys Plasmas \textbf{15}(2008)], where Riemann curvature of astrophysical plasmas loops, is responsible for unstretching by plasma flow, one obtains a torus dynamos twisted flows  slow dynamo such as in Moebius strip dynamo, considered by Shukurov, Stepanov and Sokoloff [Phys. Rev. \textbf{E 78},025301,(2008)] to modelling Perm dynamo torus of liquid sodium. Diffusion and advection (stretching), which are competing effects for dynamo action, lead us to show plasma resistivity proportionality to Riemann curvature (folding). Shukurov et al, showed that in Ponomarenko dynamo torus model a broader channel produces a better dynamo. These results agree with Schekochihin et al [Phys Rev \textbf{E} (2002)] where random magnetic structures are strengthen by curvature inverse. Analysis of spectrum of chaotic fast dynamos, shows curvature acts as a damping, since growth rate is inversely proportional to Riemann curvature. Comparison with general relativistic MHD dynamo equation, shows that the Ricci tensor, which is a contraction of the Riemann tensor also appears in the diffusion term. Curved plasma loop yields a  ${{R^{1}}_{212}}|_{\textbf{Plasma}}\approx{5.6{\times}10^{-19}m^{-2}}$. Perm torus has astronger one. Slow dynamos are favoured in laboratories rather than in plasma loops. Curvature-stretching flux rope dynamo coupling energy, coincides with minimum twist energy ${\epsilon}_{\textbf{twist}}\approx{10^{30}TeV}$ stored in flux ropes. Torus flux tubes around black-holes remain in the order of $2MeV$ and GBR are around $10^{52}TeV$. ${R^{1}}_{212}$ is negative, and inflexionary flux tubes dynamos may be responsible for this CME mechanism. Resonant UHF frequencies of $100GHz$ are found in curved-twisted loops. This is close to plasma Tokamak $91GHz$ result.{\bf PACS numbers:\hfill\parbox[t]{13.5cm}{47.65.Md, 96.60.pc}}}
\newpage \section{Introduction}
 Dynamo action in astrophysical plasmas \cite{1} concerning dynamo twisted magnetic flux tubes or flux ropes \cite{2} or liquid sodium dynamo laboratories \cite{3} is a dispute between advection-stretching terms in the self-induction magnetic equation and diffusion terms. While the first ones, enhance dynamo action, diffusion acts as to oppose fast dynamo action and in a certain sense try to diffuse magnetic field lines and spoil concentration. An important factor which gave rise recently to an anti-fast dynamo theorem proposed by Vishik \cite{4} and recently generalized by Garcia de Andrade \cite{5} for kinematic screw (helical) dynamo action in plasmas, is that unstretching of magnetic field lines or non-stretched lines at all leads in general to slow dynamos or to a decay in the magnetic field when diffusion is strong or for low magnetic Reynolds numbers $Rm$. \newline
 This problem is important in the sense that in many interesting dynamo action problems in laboratory, such as in Perm turbulent liquid sodium dynamo experiment \cite{3}, analytical solutions of the induction or dynamo equation are really difficult and can only be realized by dynamo simulation experiments with the aid of computer codes. They also obtained analytical solutions in the dynamo Moebius channels that indicates that slow dynamos are favoured. Actually by using broader torus channels they obtained still better dynamos. In this paper one obtains by simply analyzing the analytical behaviour of those advection-stretching and diffusion terms in a simple flux tube model which serves the purposes of distinct physical models of importance to dynamo theory. In the first it is analyzed the behaviour of the diffusion and stretching terms in the thin and twisted magnetic flux tubes under the influence of Riemannian curvature. In this case it is shown that Riemann curvature two-dimensional tensor (Gaussian curvature), which is composed of the product of the two principal directions of the surface, does not enhance fast dynamo action. This is due to the fact that the toroidal magnetic field lines in the stretching term, are shown to be divided by the linear Riemann curvature component term ${R^{1}}_{212}$ term, while the diffusion term is multiplied or enhanced by the same curvature term. This lead us to conclude that not only thin extremely curved solar flux tubes, tend to support slow dynamos but if one considers thick flux tubes, the same reasoning can be applied to the torus liquid sodium dynamo experiment where another example of twisted dynamo flow, the Moebius strip dynamo flow has been recently obtained by Shukurov et al \cite{6}. One of main differences to their analysis is that here one considers that the cross-section is constant while in their computations an elliptical cross-section is found. When the dynamo flow is constrained \cite{7} by cylindrical walls like in Riga and Karlshure dynamo experiments \cite{8}, the Riemann space is Riemann-flat or vanishes and in this sense dynamo action shall probably be a fast dynamo action. This flow of a cylindrical periodic tube conducting wall, has been
 investigated recently by Dobler, Frick and Stepanov \cite{3}. Riemann flat toroidal space of the dynamo experiment, is embbeded in the Euclidean three-dimensional $\textbf{E}^{3}$ space of laboratory where a dynamo action is obtained by a sudden braking of torus rotation inducing dynamo flow turbulence. The mechanism of stretching the magnetic field lines to generate dynamo action \cite{9} first appears in the literature in a paper by Arnold, Zeldovich, Ruzmaikin and Sokoloff \cite{10} in the case of a uniform stretching. This paper represented the first exact analytical solution of fast dynamos. Another example of the role of Riemann curvature tensor in dynamos, was a stretch-twist and fold fast dynamo action \cite{11} in conformal Riemannian manifolds. In the torus case the overall Gaussian curvature vanishes, nevertheless the lower internal part of the torus possesses a negative curvature while the external part possesses a positive curvature. By the way in inflexional solar flux tubes, the Riemann curvature is negative and the dynamo must be slow in this region. Another, important aspect of the application of Riemann geometry in plasma astrophysics of dynamos, is that the kinetic flow helicity responsible for the ${\alpha}{\omega}$-dynamo, where ${\omega}$ is the differential rotation of astrophysical dynamos \cite{12}, can be shown to be computed in terms of the Riemann curvature, so twist and helicity seems to be enhanced by Riemann curvature which in a certain sense may favour fast dynamos in these kind of ${\alpha}$-helical dynamos in astrophysical settings. This dynamo has been tested experimentally on a set-up in New Mexico university \cite{13}. Two important points cannot not be left aside in this introduction. The first is that the random mean field MHD cannot always gives some good results in dynamo theory, as been recently pointed out by Brandenburg \cite{2}. Therefore this is the main reason by which one uses here kinematic dynamos non-mean field dynamo theory \cite{14}. Note that other fields such as gravitational radiation can be connected with dynamo action torus around the black holes. This sort of dynamo torus around the black-hole accretion disks are considered by Van Putten \cite{15}.\newline
 \hspace{4.0cm}Another interesting point is that Schekochihin, Cowley and Maron \cite{16}, showed that under compression, or folding which is equivalent here to our Riemann curvature effect, the magnetic field lines strength in inversely proportional to line curvature. This result seems to be the counterpart here to our results here when line curvature is substituted by non-random metric surfaces endowed with Riemann two-dimensional curvature.  \newline
 \hspace{4.0cm}The paper is organised as follows: Section II presents computations regarding the stretching and diffusion in two-dimensional manifolds. Section III, addresses the issue that even in chaotic diffusion-free dynamos, the speed of dynamo action is damped by curvature. In Section III one also shows that the curvature-stretching coupling energies stored in plasma dynamo flux ropes are of the order of the CMEs (coronal mass ejection). Section IV presents conclusions.
 \newpage
\section{Stretching-curvature effects in slow dynamos in flux tubes and torus devices}
Let us now considering the equation for the self-induced MHD dynamo, where a magnetic flow field $\textbf{B}$ is given in curved Riemannian tube coordinates $({r},{\theta}_{R},s)$ by the rule
\begin{equation}
\textbf{B}= B_{\theta}\textbf{e}_{\theta}+B_{s}\textbf{t}\label{1}
\end{equation}
where $(\textbf{e}_{r},\textbf{e}_{\theta},\textbf{t})$ is the tubular frame attached to the surface of the magnetic flux tube. Thus by considering the twisted flow in magnetic flux tube Riemann metric
\begin{equation}
dl^{2}= dr^{2}+r^{2}d{{\theta}_{R}}^{2}+K^{2}(r,s)ds^{2} \label{2}
\end{equation}
By taking $K(r,s)=(1-{\kappa}_{1}r_{0}\cos{\theta}):=1$ , where  ${\kappa}_{1}$
is the Frenet external curvature and the twist transformation angle
\begin{equation}
{\theta}(s):={\theta}_{R}-\int{{\tau}(s)ds} \label{3}
\end{equation}
one obtains, the Riemannian line element of the thin flux tube, as
\begin{equation}
dl^{2}= dr^{2}+r^{2}d{{\theta}_{R}}^{2}+ds^{2} \label{4}
\end{equation}
Throughout the paper one shall consider the thin flux tubes approximation, which is suitable for solar dynamo flux tubes. Though the solar dynamos in general consider the ${\alpha}{\Omega}$-dynamo including differential rotation ${\Omega}$ to stretching the tube instead of the ${\alpha}^{2}$-dynamo consider here. Of course, ${\alpha}^{2}$-dynamos can be used in solar turbulent flows. Since in solar loops or thin magnetic twisted flux tube, upflows are turbulent while, the downflows are laminar, one may not need use here the framework of more complicated turbulent magnetic dynamo flows, and simpler kinetic non-turbulent phase of the dynamos can be used to simplify matters. Therefore within this approximation, one may use the following form of the magnetic induction equation
\begin{equation}
{\partial}_{t}\textbf{B}=
{\nabla}{\times}(\textbf{v}{\times}\textbf{B})+{\eta}{\Delta}\textbf{B}\label{5}
\end{equation}
where the gradient in Riemannian curvilinear thin flux tube coordinates is
\begin{equation}
{\nabla}=\textbf{e}_{r}{\partial}_{r}+[\textbf{t}-\frac{{\kappa}_{0}}{{\kappa}_{1}}\textbf{e}_{\theta}]{\partial}_{s}
\label{6}
\end{equation}
Note that ${\kappa}_{0}=\frac{1}{r_{0}}$ is the internal Frenet scalar curvature of the flux tube, while $r_{0}$ is the radius of the internal circular cross-section, of the tube. Here also ${\kappa}_{1}=\frac{1}{r_{1}}$ is the corresponding external counterpart of the toroidal flux tube. The Frenet tube frame attached to the magnetic flux tube axis possesses the following evolution equations
\begin{equation}
\frac{d\textbf{t}}{ds}={\kappa}_{1}(s)\textbf{n}
\label{7}
\end{equation}
\begin{equation}
\frac{d\textbf{n}}{ds}=-{\kappa}_{1}(s)\textbf{t}+{\tau}_{1}\textbf{b}
\label{8}
\end{equation}
\begin{equation}
\frac{d\textbf{b}}{ds}=-{\tau}_{1}(s)\textbf{n}
\label{9}
\end{equation}
The relation between the Frenet basis $(\textbf{t},\textbf{n},\textbf{b})$ and the other curvilinear frame is given by
\begin{equation}
{\partial}_{s}\textbf{e}_{\theta}=-{\tau}_{0}\sin{\theta}\textbf{t}\label{10}
\end{equation}
and
\begin{equation}
\textbf{e}_{\theta}=-\sin{\theta}\textbf{n}+\cos{\theta}\textbf{b}\label{11}
\end{equation}
\begin{equation}
\textbf{e}_{r}=\cos{\theta}\textbf{n}+\sin{\theta}\textbf{b}\label{12}
\end{equation}
By considering that the Riemann curvature tensor component of the flux tube surface is given by the product of the internal and external curvature of the tube as
\begin{equation}
{R^{1}}_{212}={\kappa}_{0}{\kappa}_{1}\label{13}
\end{equation}
with these mathematical tools at hand, now one can obtain the following expression for the stretching term in the induction equation
\begin{equation}
(\textbf{B}.{\nabla})\textbf{v}=[B_{s}-\frac{{\kappa}_{0}}{{\kappa}_{1}}B_{\theta}]{\partial}_{s}[v_{s}\textbf{t}+v_{\theta}\textbf{e}_{\theta}]
\label{14}
\end{equation}
which after some algebra yields
\begin{equation}
(\textbf{B}.{\nabla})\textbf{v}=[B_{s}-\frac{{\kappa}_{0}}{{\kappa}_{1}}B_{\theta}]{\omega}
{\kappa}_{1}[\textbf{n}+{\kappa}_{1}sin{\theta}\textbf{t}]({R^{1}}_{212})^{-1}
\label{15}
\end{equation}
where the frequency ${\omega}_{\theta}={\omega}$ is resonant state of poloidal and toroidal frequencies ${\omega}$ and ${\omega}_{\theta}$ is given by
\begin{equation}
v_{s}={\omega}r_{1}=\frac{{\omega}}{{\kappa}_{1}}
\label{16}
\end{equation}
\begin{equation}
v_{\theta}={\omega}_{\theta}r_{0}=\frac{{\omega}_{\theta}}{{\kappa}_{0}}
\label{17}
\end{equation}
Note that one is considering here that the dynamo flow is incompressible
\begin{equation}
{\nabla}.\textbf{v}=0
\label{18}
\end{equation}
\begin{equation}
{\nabla}.\textbf{B}=0
\label{19}
\end{equation}
where this can be explicitly given by
\begin{equation}
{\partial}_{s}{v}_{\theta}=\frac{v_{\theta}{\kappa}_{1}}{{\kappa}_{0}}sin{\theta}
\label{20}
\end{equation}
followed by the analogous equation for the magnetic field $\textbf{B}$ as
\begin{equation}
{\partial}_{s}{B}_{\theta}=\frac{B_{\theta}{\kappa}_{1}}{{\kappa}_{0}}sin{\theta}
\label{21}
\end{equation}
In terms of the Riemann curvature ${R^{1}}_{212}={R^{\theta}}_{s{\theta}s}$, these expressions become
\begin{equation}
{\partial}_{s}{v}_{\theta}=\frac{v_{\theta}{R^{1}}_{212}}{{{\kappa}_{0}}^{2}}sin{\theta}
\label{22}
\end{equation}
\begin{equation}
{\partial}_{s}{B}_{\theta}=\frac{B_{\theta}{R^{1}}_{212}}{{{\kappa}_{0}}^{2}}sin{\theta}
\label{23}
\end{equation}
Note that from expression (\ref{15}) the relation between the stretching term and Riemann curvature can be expressed as
\begin{equation}
(\textbf{B}.{\nabla})\textbf{v}\approx{{\omega}
{\kappa}_{1}[\textbf{n}+{\kappa}_{1}sin{\theta}\textbf{t}]({R^{1}}_{212})}^{-1}
\label{24}
\end{equation}
From this expression one notes that as the Riemann tensor ${R^{1}}_{212}\rightarrow{0}$ the stretching terms grows and fast dynamo action is enhanced, in this case as one shall shortly see, the diffusion term vanishes, since it is proportional to the Riemann curvature tensor component. Therefore in this case one would be left with the case of the non-diffusive ideal plasma of chaotic dynamos. This case shall be examined in the next section. Let us then compute the magnetic diffusion
\begin{equation}
|{\eta}{\Delta}\textbf{B}|=|{\eta}{\nabla}^{2}\textbf{B}|=|{\eta}[1-\frac{{{\kappa}_{0}}^{2}}
{{{\kappa}_{1}}^{2}}]
{{\partial}_{s}}^{2}[B_{\theta}\textbf{e}_{\theta}+B_{s}\textbf{t}]|
\approx{{\eta}|({R^{1}}_{212})^{-2}|}
\label{25}
\end{equation}
where ${\eta}$ is the diffusion constant. One notes that the diffusion term has a nonlinear contribution to the inverse of a nonlinear Riemann curvature tensor. To end this section let us compute the helicity ${\alpha}=-\frac{{\tau}^{*}}{2l}[\textbf{v}.{\nabla}{\times}\textbf{v}]$ which is
\begin{equation}
{\alpha}=-\frac{{\tau}^{*}}{2l}={\kappa}_{0}sin{\theta}\approx{-{\kappa}_{1}{\kappa}_{0}}
\approx{-({R^{1}}_{212})L}
\label{26}
\end{equation}
where one has used the approximation of the small ${\theta}$ angles, where ${\theta}\approx{{\theta}_{R}-{\kappa}_{1}L}$ where one has used the helical hypothesis ${\kappa}_{1}={\tau}_{1}$, where ${\tau}_{1}$ is the Frenet scalar torsion, and L is the  solar loop length. In the next section one shall show use the following ansatz
\begin{equation}
\textbf{B}=e^{{\lambda}t}\textbf{B}_{0}(s)
\label{27}
\end{equation}
Thus one obtains
\begin{equation}
{\partial}_{t}\textbf{B}={\lambda}\textbf{B}+{\gamma}[{\kappa}_{1}(B_{s}-{R^{1}}_{212}
sin{2}{\theta})\textbf{n}-B_{\theta}{\kappa}_{1}sin{{\theta}}\textbf{t}+B_{\theta}{{\kappa}_{1}}^{2}
cos{{\theta}}\textbf{b}]
\label{28}
\end{equation}
where ${\gamma}={\omega}_{\theta}-{\omega}$ vanishes due to the resonant hypothesis. This results the last expression to
\begin{equation}
{\partial}_{t}\textbf{B}={\lambda}\textbf{B}
\label{29}
\end{equation}
Thus the resonant hypothesis, tremendously simplify the future computations.
\newpage
\section{Chaotic fast dynamo and curvature-coupling CMEs energies}
Torus geometry has been proved very useful in several areas of galactic astronomy and even general relativistic jets \cite{15}. Though one has not address this problem here is possible to predict poloidal observed magnetic fields up to $10^{4}G$ from computations of the model discussed in last section. Since the Riemann curvature is small and as shown in last section diffusion terms is directly proportional to curvature, here one shall neglect diffusion to compute a dynamo analytic and complete solution of the induction equation in the case of chaotic dynamo on a torus or plasma loops. Let us first compute the Riemann curvature for the Riemann curvature-stretching coupling (RS), for a non-flare loop \cite{17}. In these coronal plasma loops the radius of the loop is given by $r_{1}\approx{10^{11}cm}$ and the thickness of the loop is given by $r_{0}\approx{10^{8}cm}$. Therefore the respective scalar curvatures are given by ${\kappa}_{1}\approx{10^{-8}cm^{-1}}$ and ${\kappa}_{0}\approx{10^{-10}cm^{-1}}$. Thus the Riemann curvature tensor is given by
\begin{equation}
{R^{1}}_{212}={\kappa}_{0}{\kappa}_{1}\approx{10^{-19}cm^{-1}}
\label{30}
\end{equation}
Thus to be able to compute the magnetic energy associated with the curvature-stretching coupling, one needs to solve the diffusion-free equation
\begin{equation}
{\partial}_{t}\textbf{B}=
{\nabla}{\times}(\textbf{v}{\times}\textbf{B})\label{31}
\end{equation}
The three scalar equations obtained for this equation in the case of plasma loops yields the following constraint on the poloidal and toroidal field of the manifold
\begin{equation}
\frac{B_{s}}{B_{\theta}}={R^{1}}_{212}{{\kappa}_{1}}^{-1}
\label{32}
\end{equation}
If one substitute ${{\kappa}_{1}}^{-1}$ by the radius $r_{1}$, one obtains an expression for the homogeneous twist \cite{18}, since as one has seen above the helical curvature coincides with torsion of the flux tubes. The other equations yields
\begin{equation}
{\lambda}+\frac{B_{\theta}}{B_{s}}\frac{(1+{\omega}{\kappa}_{1})}{{\kappa}_{1}}=0
\label{33}
\end{equation}
The last two equations yield
\begin{equation}
{\lambda}=-{{\kappa}_{0}}^{-1}[1+{R^{1}}_{212}{\omega}{{\kappa}_{0}}^{-1}]
\label{34}
\end{equation}
Note that the only way that this equation for the growth rate ${\lambda}$ yields the fast dynamo limit
\begin{equation}
lim_{{\eta}\rightarrow{0}}{\lambda}\ge{0}
\label{35}
\end{equation}
is that the Riemann curvature ${R^{1}}_{212}$ be negative. This is in agreement with a previous result of Chicone et al \cite{19} that the only way to have a two dimensional compact manifold of Riemann constant curvature is that this be negative. If one calls ${R^{1}}_{212}=-{\beta}^{2}$
\begin{equation}
{\beta}^{2}{\omega}\ge{{\kappa}_{0}}
\label{36}
\end{equation}
or the constrain on the Riemann curvature
\begin{equation}
|{R^{1}}_{212}|{\omega}\ge{{\kappa}_{0}}
\label{37}
\end{equation}
This expression places a lower bound on the resonant frequency ${\omega}$ from the plasma loop data as
\begin{equation}
{\omega}\ge{{\kappa}_{0}|{R^{1}}_{212}|^{-1}}
\label{38}
\end{equation}
or
\begin{equation}
{\omega}\ge{|{\kappa}_{1}|^{-1}}\approx{10^{11}Hz}
\label{39}
\end{equation}
This is a ultra-high frequency (UHF). Let us now consider the equation (\ref{34}) to compute the magnetic energy associated with plasma loop fast dynamos. This yields an expression
\begin{equation}
{\epsilon}=\frac{1}{8{\pi}}\int{B^{2}dV}\approx{{B_{\theta}}^{2}|{R^{1}}_{212}|^{-1}}
\label{40}
\end{equation}
for poloidal fields of $10G$ yields a Riemann-stretching coupling energy, one obtains the value ${\epsilon}_{\textbf{RS}}\approx{5.6{\times}10^{30}TeV}$ which is of the same order of the minimum twist energy of CMEs computed by Moore \cite{20}, in the range $10^{30}-10^{32}ergs$. Though this values range in the order of many CERN energies, huge energies in the universe as for example in GBR supernovas can be estimated as high as $10^{52}ergs$. Torus flux tubes around magnetospheres around general relativistic black-holes are in the scale of some MeV. Since here the Riemann curvature is negative one is lead to conclude that the inflexionary plasma loops are present in this chaotic dynamo which pave the way to CMEs.
\newpage
\section{Conclusions}
Recently Kleeorin et al \cite{21} have investigated the random magnetic field evolution of magnetic field based on a mean-field magnetohydrodynamics, in two distinct type of flows called Roberts flow and Arnold-Beltrami-Childress (ABC) flows. They showed that the slow Roberts dynamo could be transformed into a fast dynamo by mean-field helicity and stretching. Here a similar transformation is consider such as an inverse of that, in the sense that now, the fast dynamos could be damped in its Riemann curvature-stretching coupling. Results obtained by this idea were already given by Schekochihin et al \cite{10} in the case of random filaments where curvature would be the mean value of the scalar curvatures of the filaments. Among many interesting physical consequences are the applications of the model here to the Perm dynamo torus. Other interesting future implications of this torus geometry can be obtained in the case of astrophysical plasma dynamo accretion disc torus magnetospheres around general relativistic black-holes. Plasma torus can be analyzed in terms of solar corona ejections. This lead us to show that chaotic fast dynamos may lead to values of the order of the twist energy of CMEs flux rope computed by Moore \cite{20}. The Perm experimental dynamo is not the only experiment where the dynamo is slow but a recently proposal by de La Torre and Burguette et al \cite{22} has also used a slow dynamo flow. Future prospect of torus dynamos include further investigation of the models of non-relativistic dynamos to apply them to the GR MHD dynamos by using equations recently proposed by Clarkson and Markllund \cite{23}. One of the most important features of the plasma curved and twisted model discussed here is that it possesses a resonant frequency of the order of the one obtained in plasma tokamak devices, for electron plasmas \cite{24}, which are also presented in plasma loops. Threfore one must conclude that the Riemann-stretching plasma loop model discussed here is fits well within experimental results produce in tokamak plasma laboratories and plasma astrophysical flux tubes. In heliotrons where torsion is present as in the twisted plasma dynamos considered here, lower frequencies of the order of 53GHz can be achieved \cite{25}.
\section{Acknowledgements} I also deeply in debt to Dmitry Sokoloff, Javier Burguette, Guenther Ruediger, Manfred Schuessler and Renzo Ricca for helpful discussions on
the subject of this paper. Financial supports from Universidade do
Estado do Rio de Janeiro (UERJ) and CNPq (Brazilian Ministry of
Science and Technology) are highly appreciated.

 \end{document}